\documentstyle[supertabular,epsf,rotate]{mn2e}
\newcommand{\ltsimeq}{\raisebox{-0.6ex}{$\,\stackrel 
        {\raisebox{-.2ex}{$\textstyle <$}}{\sim}\,$}} 
\newcommand{\gtsimeq}{\raisebox{-0.6ex}{$\,\stackrel 
        {\raisebox{-.2ex}{$\textstyle >$}}{\sim}\,$}} 

\input epsf.sty

\begin{document}

\title[The influence of powerful jets on galaxy evolution]
{Evidence that powerful radio jets have a profound
influence on the evolution of galaxies} 

\author[Rawlings \& Jarvis]{
Steve Rawlings\footnotemark and 
Matt J.\ Jarvis\\ 
Astrophysics, Department of Physics, Denys Wilkinson Building, Keble Road,
Oxford, OX1 3RH, UK 
}

\maketitle

\begin{abstract}
\noindent
The relationships between supermassive black holes and the properties of their 
associated dark-matter halos imply that outflows from accreting black holes 
provide a feedback mechanism regulating galaxy formation.
Accreting black holes with weak or undetectable radio jets (radio-quiet quasars) 
outnumber those with powerful jets (radio-loud quasars) by a factor $\sim 10-100$, 
so powerful-jet outflows are often neglected. However, whenever powerful jets are triggered,
there is a dramatic (factor $\gtsimeq 100$) step-function increase in the
efficiency of feedback. We use a feedback model, together
with the measured space density of flat-spectrum radio-loud quasars, to show 
that a powerful-jet episode probably occurred in every protocluster in the 
Universe. Before jet triggering, there was time for gravitational collapse to 
create many ($\sim 10-100$) surrounding protogalaxies massive enough to host
radio-quiet quasars. After triggering, the powerful jet pushes back 
and heats ionized gas so that it cannot fall onto these protogalaxies 
and cool. Once neutral/molecular gas reservoirs become exhausted,
there is a synchronized shut down in both star-formation and
black-hole activity throughout the protocluster. These considerations imply that
radio-loud quasars have a profound influence on the evolution of all
the galaxies seen in clusters today. 
\end{abstract}


\footnotetext{Email: s.rawlings1@physics.ox.ac.uk}

\section{Introduction}

Powerful-radio-jet activity is regarded as a useful tracer of
large-scale structure in the distant Universe (e.g.\ Miley et al.\ 2004), but with 
the exception of a few studies (e.g.\ Gopal-Krishna \& Wiita 2001;
Rawlings 2003), its influence on galaxy formation is
typically ignored. This is because, even in the young Universe where 
supermassive black holes have high accretion rates
and are visible as quasars, only a few per cent of accreting 
black holes develop powerful radio jets (e.g.\ Goldschmidt et al.\ 1999). 
It is now accepted, however, that quasars do 
play a key r\^{o}le in galaxy formation because of the 
existence of remarkably tight correlations
between the masses of black holes and the properties of their
associated dark-matter halos such as velocity dispersion (e.g.\ Ferrarese 2002).

Such correlations are most easily understood in terms of `feedback 
models' (Silk \& Rees 1998; Fabian 1999)  
in which the mechanical power emerging from radio-quiet accreting 
black holes injects energy into the gaseous component of the young galaxy
over and above that which it has acquired by gravitational 
collapse. In Sec.~\ref{sec:feedback} we adopt a feedback model for
radio-loud accreting black holes and in Sec.~\ref{sec:evolution} we use this
model, together with hierarchical clustering theory, to compare the 
predicted cosmic evolution in the comoving space density of radio sources
with the evolution observed. We reach some new conclusions concerning the 
cosmological importance of radio sources in Sec.~\ref{sec:conclusions}.

The convention for spectral index $\alpha$ is that flux density 
$S_{\nu} \propto \nu^{-\alpha}$, where $\nu$ is the observing frequency, and
the radio luminosity function (RLF), the comoving space density of sources
per (base 10) logarithmic interval of 1.4-GHz radio luminosity $L_{1.4}$, is assumed
proportional to $L_{1.4}^{- \beta}$.
We assume throughout a low-density, $\Lambda$-dominated 
Universe in which $h = H_{0} / (100 ~ \rm km ~ s^{-1} ~ Mpc^{-1}) = 0.7$; 
$\Omega_{\rm m} = 0.3$; $\Omega_{\Lambda} = 0.7$; $\Omega_{b} = 0.04$,
$\sigma_{8 ~ h^{-1}} = 0.9$, and $n_{\rm scalar} = 1$.  

\section{Feedback due to powerful radio jets}
\label{sec:feedback}

Here, we adopt a form of feedback model in which powerful
(radio-loud quasar) jets deliver 
some fraction $f_{\rm effic}$ of their
mechanical power $Q$ to ionized gas which, prior to the powerful-jet
episode, is bound to a number $N_{\rm halo}$ of
dark-matter-dominated
halos (one hosting the radio-jet-producing central engine, 
the others in the surrounding protocluster), each
of mass $M_{\rm halo}$ and velocity dispersion $\sigma$. 
We assume this injection of energy occurs over a timescale 
$t_{\rm life}$, and that the energy is deposited as thermal energy in the 
gas (with negligible radiative losses; see Rawlings 2003), and that each halo has just 
virialized at redshift $z$, so that $\sigma \propto 
M_{\rm halo}^{\frac{1}{3}} ~ (1+z)^{\frac{1}{2}}$ (Somerville \& Primack 1999).
We consider the critical point at which sufficient mechanical energy is 
delivered to the gas so that it just becomes unbound from 
each of the $N_{\rm halo}$ surrounding halos. This yields a scaling relation

\small
\begin{eqnarray}
\label{eq:one}
\left( \frac{f_{\rm effic}}{0.5} \right) 
\left( \frac{Q}{1.25 \times 10^{40} ~ \rm W} \right) 
\left( \frac{t_{\rm life}}{4 \times 10^{7} ~ \rm yr} \right) \nonumber \\
\sim
\left( \frac{N_{\rm halo}}{100} \right)
\left( \frac{f_{\rm gas}}{0.13} \right)
\left( \frac{M_{\rm halo}}{5 \times 10^{12} ~ \rm M_{\odot}} \right)
\left( \frac{\sigma}{290 ~ \rm km ~ s^{-1}} \right)^{2} \nonumber \\ 
\sim k_{1} \left( \frac{N_{\rm halo}}{100} \right) 
\left( \frac{\sigma}{290 ~ \rm km ~ s^{-1}} \right)^{5} \\ \nonumber
\end{eqnarray}
\normalsize

\noindent
where $f_{\rm gas}$ is the fraction of the total (dark-matter-dominated) 
mass in gas, and $k_{1}$ is a constant of order unity. The choices of
normalizing constants for each variable will be explained in 
Sec.~\ref{sec:evolution}.

It is clear from Equation~1 that powerful-jet activity can remove gas 
not only from a single host galaxy, but also from a large number ($N_{\rm halo}
\sim 100$) of surrounding 
galaxy-sized dark-matter halos. In the local Universe, powerful-jet activity is 
confined to black holes triggered in rich clusters of galaxies (e.g.\ Cygnus A and 3C~295),
and although X-ray observations provide ample evidence of ionized gas being pushed back 
and heated by such radio sources (e.g.\ Smith et al.\ 2002), the cluster potential
wells are sufficiently deep ($\sigma \sim 1000 ~ \rm km ~ s^{-1}$) that the gas
remains in a gravitationally-bound intracluster medium. 
This will not be true in similar systems at much earlier times because hierarchical 
structure formation (e.g.\ Press \& Schechter 1974; Percival, Miller \& Peacock 2000) 
demands that, in the young Universe, the clusters of galaxies seen today were 
gravitationally-unbound protoclusters, collections of protogalaxies lacking
any deep extended potential well, and hence any intracluster medium.

\section{Comparison of predicted and observed cosmic evolution of radio sources}
\label{sec:evolution}

The hierarchical clustering of dark-matter halos and the background 
cosmology (Spergel et al.\ 2003) are now so well understood that it is possible to use 
the feedback model of Sec.~\ref{sec:feedback}
to make a firm prediction for the number of powerful-jet episodes triggered 
per comoving volume per unit cosmic time, modulo just one key uncertainty, the fraction of 
newly-created halos that give rise to powerful-jet activity. Powerful-jet activity 
in the local Universe gives vital clues to the critical
features of the halos: such jets emerge only from massive elliptical galaxies, with a 
relatively narrow spread in black hole mass $M_{\rm BH} \sim 10^{9 \pm 0.5} ~ \rm M_{\odot}$, 
corresponding to halo velocity dispersions $\sigma \sim 290 \pm 60 ~ \rm km ~ s^{-1}$ 
(McLure et al.\ 2004). We introduce a factor $f_{\rm halo}$ to allow for the fact that a
halo within this range of $\sigma$ is a necessary, but not sufficient,
condition for powerful-jet activity.

Some choices of normalizing constants in Equation~\ref{eq:one} follow
from the assumed close mapping between black hole and dark-matter properties: 
$Q = 1.25 \times 10^{40} ~ \rm W$ corresponds to
the most powerful jets observed (Rawlings \& Saunders 1991), 
and is the Eddington luminosity of an 
$M_{\rm BH} \sim 10^{9} ~ \rm M_{\odot}$ black hole; and
$M_{\rm halo} \sim 5 \times 10^{12} ~ \rm M_{\odot}$ is the mass of a dark-matter halo
with $\sigma = 290 ~ \rm km ~ s^{-1}$, collapsing at $z \sim 2.5$ (Somerville \& Primack 1999).
Other choices were motivated as follows: theory demands $f_{\rm effic} \sim 0.5$ 
(e.g.\ Bicknell et al.\ 1997); $f_{\rm gas} = 0.13$ corresponds to the ratio of baryons to
dark matter, assuming that baryons in forms other than hot gas can 
be neglected; and
$t_{\rm life} = 4 \times 10^{7} ~ \rm yr$ is (for an assumed quasar 
accretion efficiency $\sim 0.1$) the mass-doubling timescale for Eddington-limited
growth of a black hole, consistent with lower-limits on radio source lifetimes derived from
the observed linear sizes of powerful radio galaxies 
(e.g.\ Kaiser, Dennett-Thorpe \& Alexander 1997; Blundell \& Rawlings 1999). 

Powerful jets are
liable only to be triggered when two supermassive black holes 
coalesce (Wilson \& Colbert 1995), and since this can only happen as the 
result of a major merger (a special class of `halo creation event'), we have used hierarchical 
structure formation theory (Percival et al.\ 2000) to 
predict how the trigger rate of powerful jets depends on cosmic epoch (Fig.~1). 
Note that episodes of powerful-jet activity are predicted to occur over a wide range
of cosmic epochs, persisting at some level throughout the later
stages of the `epoch of reionization' during which a partially ionized Universe 
at $z \sim 15$ (Spergel et al.\ 2003) becomes fully ionized by $z \sim 6$
(Becker et al.\ 2001).

\begin{figure*}
\begin{center}
\setlength{\unitlength}{1mm}
\begin{picture}(150,120)
\put(-30,-20){\includegraphics{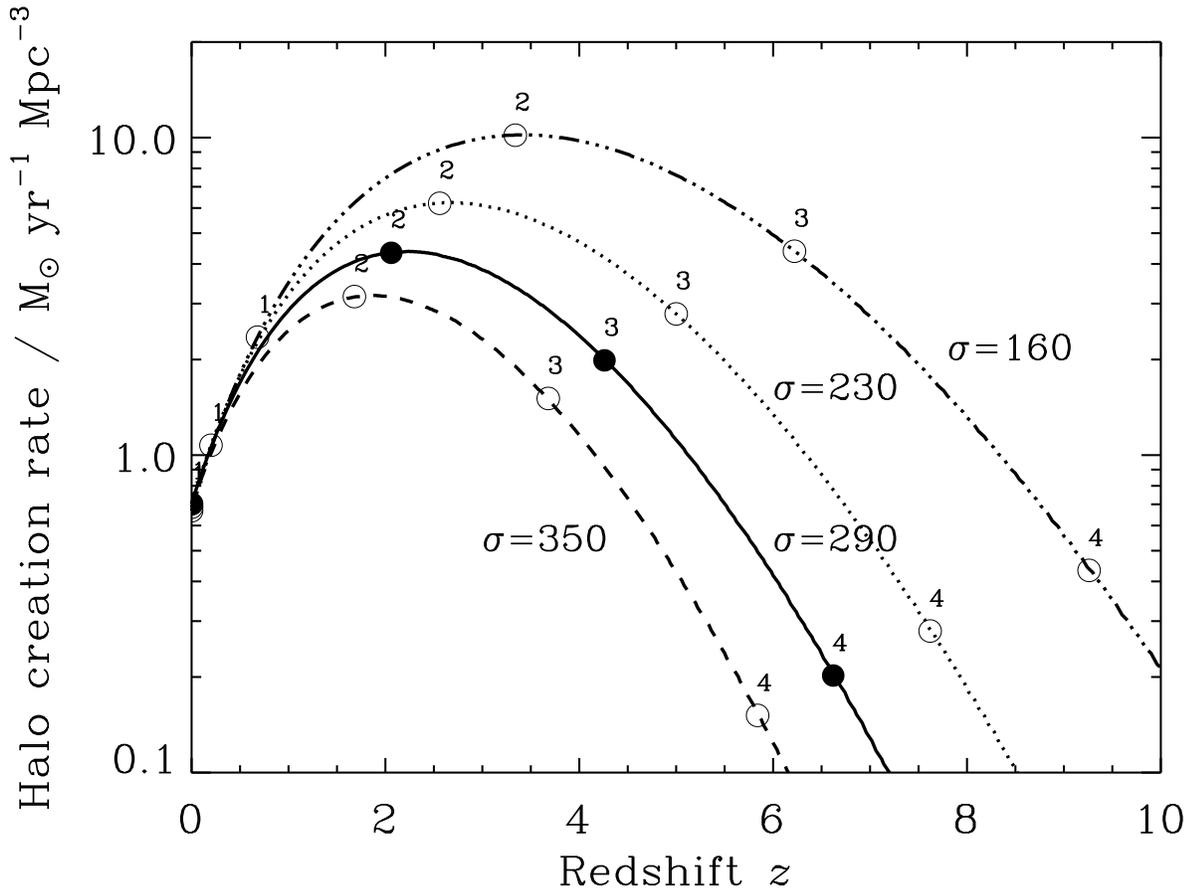}}
\end{picture}
\end{center}
{\caption[junk]{\label{fig:fig1} 
The creation rate of dark-matter halos 
as a function of redshift $z$ for various halo velocity dispersions $\sigma$
(in units of $\rm km ~ s^{-1}$): $\sigma=160$, dot-dashed line; $\sigma=230$, dotted line;
$\sigma=290$, solid line;
$\sigma=350$, dashed line. The points corresponding to $\nu = 1,2,3$ and $4$ 
for each $\sigma$ are marked, where $\nu$ is the density threshold
for collapse in units of the r.m.s. density fluctuation $\sigma_{\rho}$: $\nu = \delta_{\rm crit} /
\sigma_{\rho} (M)$, where $\delta_{\rm crit} \approx 1.7$; mass $M$ is related to
a sharp $k-$space filter (used to smooth the density field of mean
value $\rho_{0}$) by $M = 6 \pi^{2} \rho_{0} k_{\rm s}^{-3}$, with $k_{\rm s}$
the cut-off value; and $\sigma$ and $M$ are related by the spherical-top-hat-collapse
model (Somerville \& Primack 199) so that $\sigma \propto
M_{\rm halo}^{1 / 3} ~ (1+z)^{1 / 2}$.
We have estimated the halo creation rates, following Percival et al.\ (2000), by: (i)
calculating, at each $z$, the value of $M$ appropriate to the target $\sigma$; (ii)
using the Press-Schechter (PS; Press \& Schechter 1974) formalism (and the 
$\sigma_{8 ~ h^{-1}}$-normalized power spectrum) to calculate $g(z) = \nu \exp{-\nu^{2}} / 2$; 
(iii) estimating the run of creation rate with $z$ as 
$g(z) \times h(z)$, where $h(z) = (1+z)^{\left[2.3 + (0.036 \times 
z) / (1+0.203 \times z)\right]}$ is
a fitting formula (kindly provided by W.\ Percival) for ${\rm d} \delta_{\rm crit} /
{\rm d} t$, where $t$ is the cosmic time; (iv) normalizing the curves 
to the $N-$body simulation data 
of Percival et al.\ (2000), and neglecting any variations of this rate with mass, over the
small range of interest at $z=0$.
Note that the interpretation of `halo creation rates' is far more 
straightforward at high $\nu$ (say, $\nu \gtsimeq 2$), where very few halos
are `destroyed' by being subsumed in larger halos, than for
$\nu \sim 1$ fluctuations (Percival et al.\ 2000).
}}
\end{figure*}

Predicting the radio emission from powerful-jet episodes is, in general, an 
extremely complicated function of properties like the time after the 
jet-triggering event and the gaseous environment, as well as 
observational choices like frequency. We therefore focus on episodes whose
observational manifestation will be flat-spectrum 
($\alpha \sim 0$) radio emission which arises
when the jets happen to be favourably oriented, i.e.\ within a `beaming angle' covering a 
sky fraction $f_{\rm beam} \sim 0.01$ (see Jarvis \& Rawlings 2000). 
Such `Doppler-boosted' emission arises from 
synchrotron-self-absorbed knots at the base of the jet, and the ratio of its luminosity to $Q$
should be fairly constant throughout the lifetime $t_{\rm life}$ of the 
radio source (and relatively insensitive to environment and redshift effects)
whereas, in contrast, contributions to $L_{1.4}$ from extended structures will,
owing to inverse-Compton cooling and other effects (e.g.\ Kaiser et al.\ 1997;
Blundell, Rawlings \& Willott 1999), typically be a strong function of 
time since the jets were triggered, environment 
and redshift. We use the creation rate $C_{290}$ for
$\sigma = 290 ~ \rm km ~ s^{-1}$ halos (from Fig.~1) to estimate,
following Efstathiou \& Rees (1988), the comoving space density $\Phi$ of
triggered flat-spectrum sources using

\small
\begin{eqnarray}
\left( \frac{\Phi}{10^{-10} ~ \rm Mpc^{-3}} \right)
\sim k_{2}
\left( \frac{f_{\rm halo}}{0.01} \right)
\left( \frac{f_{\rm beam}}{0.01} \right)
\left( \frac{f_{\rm RLF}}{10^{-1.75}} \right) \nonumber \\
\times \left( \frac{C_{290}}{4 ~ \rm M_{\odot} ~ yr^{-1} ~ Mpc^{-3}} \right) 
\left( \frac{t_{\rm life}}{4 \times 10^{7} ~ \rm yr} \right) \nonumber \\ 
\times
\left( \frac{M_{\rm halo}}{5 \times 10^{12} ~ \rm M_{\odot}} \right)^{-1},\\ \nonumber
\end{eqnarray}
\normalsize

\noindent
where $k_{2}$ is a constant of order unity (incorporating an assumption that there is
$\sim 1$ triggering event per halo as it evolves through an $\sim 1$-dex spread in $M_{\rm halo}$), and
$f_{\rm RLF}$ (explained fully in the caption to Fig.~2) ensures 
that $\Phi$ is integrated over only the top dex of the flat-spectrum RLF. 
This is the regime (Fig.~ 2) in which the
high-redshift space density of flat-spectrum 
quasars is observationally constrained (Jarvis \& Rawlings 2000).

\begin{figure*}
\begin{center}
\setlength{\unitlength}{1mm}
\begin{picture}(150,120)
\put(-30,-20){\includegraphics{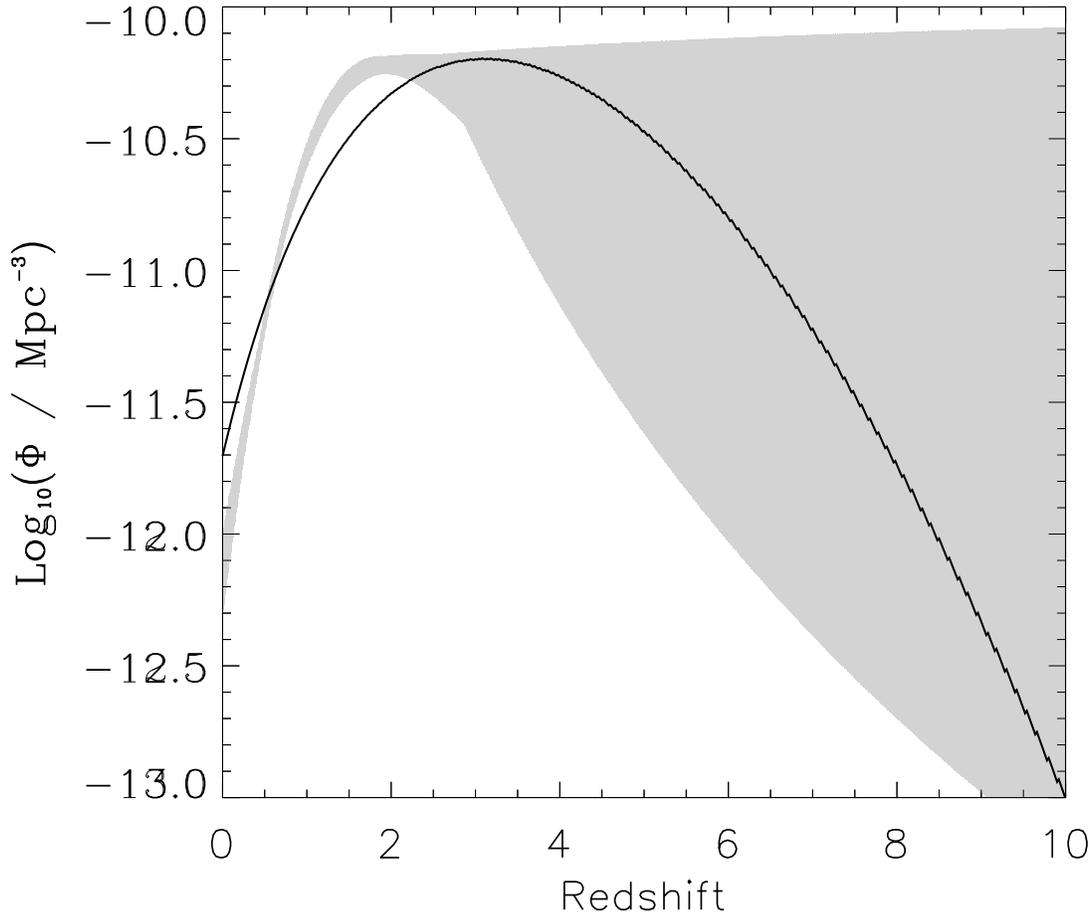}}
\end{picture}
\end{center}
{\caption[junk]{\label{fig:fig2}
Comparison of the predicted comoving space density $\Phi$ of flat-spectrum 
radio-loud quasars (solid line, calculated using Equation~2) with measured
constraints on $\Phi$ from existing surveys (shaded regions,
90 per cent confidence regions), from Jarvis \& Rawlings (2000) and
Dunlop \& Peacock (1990).
The value of $f_{\rm halo}=0.01$ was set to obtain rough agreement in the
relative normalizations.
The value of $\Phi$ is integrated only over the top dex of the RLF (as in Jarvis \& 
Rawlings 2000), 
requiring the introduction of a factor $f_{\rm RLF}$ to account for the 
lower-$L_{1.4}$ population. We adopt a scaling $L_{1.4} \propto M_{\rm BH}^2$ 
(Lacy et al.\ 2001), so that the one-dex spread in black hole masses (McLure et al.\ 2004)
maps onto a two dex spread in $L_{1.4}$. Adopting $\beta = 1.75$ (Jarvis \& Rawlings 2000), 
and going one dex further down the RLF, implies $f_{\rm RLF} \sim 10^{-1.75}$.
Note that the model and data diverge significantly at low redshift ($z \ltsimeq 2$).
This is expected because it is well known that any simple Press-Schechter-based
formalism fails to fully explain the dramatic drop in quasar activity at low redshifts.
There are several reasons for this (e.g.\ Haehnelt \& Rees, 1993) 
but all linked to two key facts (i) that 
Press-Schechter theory does not properly account for sub-halos that become
part of larger collapsed systems, e.g.\ galaxies in virialized clusters at
low redshift; and (ii) that mergers of baryonic systems like galaxies 
get strongly suppressed once they begin to inhabit larger collapsed systems in
which the velocity dispersion $\sigma$ greatly exceeds the internal velocity
dispersions of the `sub-halo' galaxies (Carlberg 1990).
}}
\end{figure*}

The crucial unknown quantity in Equation~2 is $f_{\rm halo}$.
We have fixed this at the value ($f_{\rm halo} \sim 0.01$) delivered by requiring roughly equal
normalizations for the predicted and measured values of $\Phi$ in Fig.~2. With
this normalization fixed, and within the considerable current uncertainties, the
observational constraints on $\Phi$ are in good 
agreement with the gradual high-redshift decline predicted by
hierarchical structure formation theories. However, the 
more interesting result is that we require $f_{\rm halo} \sim 0.01$,
implying that only $\sim 1$ in 100 triggering events (in the relevant halo 
velocity dispersion range) generate powerful-jet episodes.
The feedback model of Equation~1 implies that each powerful-jet episode influences, in
the young Universe,
$N_{\rm halo} \sim 100$ halos. By
considering the conditions before and after one of these episodes, 
we will argue in Sec.~\ref{sec:conclusions} that the relationship 

\small
\begin{eqnarray}
N_{\rm halo} \sim \frac{1}{f_{\rm halo}} \sim 100,
\end{eqnarray}
\normalsize

\noindent
established here using physical arguments based around Equations~1 \& 2
and a measurement of the flat-spectrum quasar RLF, 
is telling us something important about the galaxy formation process.

\section{Concluding remarks}
\label{sec:conclusions}
In the young Universe, `high-peak bias' (Kaiser 1984) will have established the
regions of space destined, by present epochs, to become clusters of galaxies. 
Each of these protoclusters will contain $\sim 100$ protogalaxies which, from Fig,~1,
will form over a wide range of epochs and which, because of
`high-peak bias', will be strongly clustered. Such protoclusters have been observed
in the form of emission-line-emitting objects and `Lyman-break' galaxies
around high-redshift radio galaxies (e.g.\ Miley et al. 2004). 
Central black holes will form in
all the high-velocity-dispersion ($\sigma > 160 ~ \rm km ~ s^{-1}$, e.g.\ 
Ridgway et al.\ 2001) halos, presumably with masses set by the feedback processes 
common in radio-quiet quasars (Silk \& Rees 1998); we see from Fig. 1 that these halos
tend to be created at much earlier epochs than the higher-velocity-dispersion ones.
Eventually, hierarchical processes will cause one pair of halos, each containing a 
supermassive black hole, to merge, creating a single $\sigma \sim 290 ~ \rm km ~ s^{-1}$
halo with a single coalesced black hole, and triggering a powerful-jet episode.
This will be a much more dramatic type of feedback event 
because the radio source injects enough energy
into its surroundings that it gravitationally unbinds ionized gas associated not only with
the host galaxy, but more widely throughout the protocluster (see also 
Nath \& Roychowdhury 2002, and refs. therein).
There has been
insufficient cosmic time for a larger dark-matter halo to form, so this 
process yields a reservoir of protocluster gas which is
not yet gravitationally bound, and is now so hot that it cannot
accrete back onto the protogalaxies. There will then be a synchronized, protocluster-wide
shut down of activity, be it circumnuclear star-formation or black-hole accretion.

For a protogalaxy $\sim 1 ~ \rm Mpc$ from the radio galaxy, fresh supplies of
neutral/molecular gas from accretion and cooling will be shut off after 
$\sim 3 \times 10^{7} ~ \rm yr$ ($\sim 10-$times the light-travel time; e.g.\
Blundell \& Rawlings 1999), and star-formation and AGN
activity will cease once the reserves of neutral gas
have been exhausted (taking $\ltsimeq 10^{8} ~ \rm yr$ if star-formation rates of $\sim 1000 ~
\rm M_{\odot} ~ yr^{-1}$ use up neutral/molecular gas
reservoirs of mass $\ltsimeq 10^{11} ~ \rm M_{\odot}$, e.g.\ Greve et al.\ 2003). 
Accounting for these time lags, 
and the `high-peak bias', it is not surprising that there seem to be significant overdensities
of both intense starbursting systems (Stevens et al.\ 2003) and X-ray-selected 
AGN (Pentericci et al.\ 2002) around high-redshift radio galaxies. There is also 
mounting evidence that the amount of gas and dust 
in the host galaxy of a powerful radio source decreases as the source expands from various 
anti-correlations between different tracers of this material and source 
sizes (Baker et al.\ 2002; Willott et al.\ 2002; Jarvis et al.\ 2003).

We conclude that powerful-jet activity represents a dramatic (factor $\gtsimeq 100$) 
step-function increase in the efficiency of feedback mechanisms believed to 
be an essential part of galaxy formation. Its influence is, however, far more
widespread as sufficient energy is delivered to the protocluster environment that 
gas can no longer cool onto the $\sim 100$ surrounding protogalaxies.
Star-formation and black hole activity throughout the protocluster is
shut down, but, accounting both for time before the jet-triggering event 
(see Fig.\ 1) and time lags after the event, there was ample opportunity for 
$\sim 10-100$ of the protogalaxies to shine as radio-quiet quasars, and to 
build up their supermassive black holes and stellar bulges. In crude terms, this 
provides a natural explanation for the relative number of radio-quiet and radio-loud quasars
at high redshift (e.g.\ Goldschmidt et al.\ 1999). The observed 
space density of flat-spectrum radio quasars (Fig.~2) predicts that
a powerful-jet episode occurs in $\sim 1$ in 100 protogalaxies, which is as
expected if each protocluster experiences $\sim 1$ such event. 

\section*{Acknowledgements}
SR is grateful to the UK PPARC for a Senior Research Fellowship, and for
financial support from the Australia Telescope National Facility. 
MJJ is supported by a PPARC PDRA. We thank Will Percival for very useful discussions.

\end{document}